\documentclass[prl,twocolumn,footinbib]{revtex4}

\begin{document}

\title{Einstein-Cartan gravity excludes extra dimensions} 
\author{Nikodem J. Pop{\l}awski}
\affiliation{Department of Physics, Indiana University, 727 E 3rd St, Bloomington, IN 47405, USA}
\email{nipoplaw@indiana.edu}
\date{\today}

\begin{abstract}
We show that the electron in the Riemann-Cartan spacetime with extra dimensions has a finite size that is much larger than the experimental upper limit on its radius.
Thus the Arkani-Hamed-Dimopoulos-Dvali and Randall-Sundrum models of the weak/Planck hierarchy in particle physics are not viable if spin produces torsion according to the Einstein-Cartan theory of gravity.
\end{abstract}

\maketitle
Newton's gravitational constant in the natural system of units ($\hbar=c=1$) is given by $G=M_{\text{Pl}}^{-2}$, where $M_{\text{Pl}}\sim10^{19}\text{GeV}$ is the Planck energy.
The Planck energy is much larger than the energy scale of the electroweak unification $M_{\text{EW}}\sim10^3\text{GeV}$.
Thus gravity is very weak compared to the other interactions.
The Arkani-Hamed-Dimopoulos-Dvali (ADD) model explains this relative weakness, which is called the hierarchy problem in particle physics, by introducing large extra dimensions \cite{ADD}.
If spacetime has $n$ extra compact spatial dimensions of radius $R$ then the gravitational potential $V(r)$ from a point mass $m$ at small distances $r\ll R$ is
\begin{equation}
V(r)\sim\frac{m}{M_{\text{Pl}(4+n)}^{n+2}r^{n+1}},
\end{equation}
while at large distances $r\gg R$ it must be equal to the usual Newtonian potential,
\begin{equation}
V(r)\sim\frac{m}{M_{\text{Pl}(4+n)}^{n+2}R^nr}.
\end{equation}
Thus the Planck energy in this $(4+n)$-dimensional spacetime, $M_{\text{Pl}(4+n)}$, is related to the Planck energy of the 4-dimensional spacetime by
\begin{equation}
M_{\text{Pl}(4+n)}^{n+2}R^n=M_{\text{Pl}}^2.
\end{equation}
In the ADD model there is only one fundamental energy scale,
\begin{equation}
M_{\text{EW}}\sim M_{\text{Pl}(4+n)}.
\end{equation}
Thus the radius $R$, which reproduces the observed $M_{\text{Pl}}$, is given by
\begin{equation}
R\sim10^{30/n-19}\text{m}.
\end{equation}
The value $n=1$ gives the size of the compactification radius $R\sim10^{11}\text{m}$, which would lead to deviations from Newtonian gravity at distances on the order of the size of the Solar System, and therefore excluded.
The value $n=2$ gives $R\sim10^{-4}\text{m}$, which is on the order of current experimental upper limits on the distance at which new macroscopic forces may exist \cite{Long}.
Since the electroweak and strong forces have been tested at electroweak scale distances, which are much smaller than $10^{-4}\text{m}$, particles are localized in the 4-dimensional spacetime and cannot propagate in the extra dimensions.
As $n\rightarrow\infty$, $R\rightarrow10^{-19}\text{m}$.

The Schwarzschild radius $r_S$ for a mass $m$ in Einstein's general relativity (GR) is on the order of $Gm$.
For the electron, $r_S\sim10^{-57}$m.
In the presence of $n$ large extra dimensions ($R\gg r_S$), it is given by \cite{MP}:
\begin{equation}
r_S\sim(G_n m)^{\frac{1}{n+1}},
\end{equation}
where the gravitational constant $G_n$ of the $(4+n)$-dimensional spacetime is related to the corresponding Planck energy by
\begin{equation}
G_n M_{\text{Pl}(4+n)}^{n+2}=1.
\end{equation}
Thus
\begin{equation}
r_S\sim(GmR^n)^{\frac{1}{n+1}}.
\label{Schw}
\end{equation}
For the electron, $n=2$ gives $r_S\sim10^{-22}$m.
As $n\rightarrow\infty$, $r_S\rightarrow10^{-19}$m.
While the Schwarzschild radius of the electron in GR is much smaller than the upper limit on the particle's radius $\sim10^{-22}\text{m}$ observed in a Penning trap \cite{rad}, the Schwarzschild radius of the electron in the ADD model is on the order of this limit, imposing strong constraints on the physically possible parameters of this model (since the size of a particle is expected to be on the order of its Schwarzschild radius \cite{Wald}).
In the nonrelativistic limit of the ADD model, the Einstein equations reduce to the Poisson equation $\triangle V=4\pi G\rho$, so for objects of sizes $r\ll R$, the mass density is
\begin{equation}
\rho\sim\frac{mR^n}{r^{n+3}}.
\end{equation}
The Planck mass $M_{\text{Pl}(4+n)}\sim10^3$GeV of the $(4+n)$-dimensional spacetime gives also the order of the theoretical minimum mass of a black hole in the ADD model.
Therefore if this model is true, the LHC, which operates at energies on the order of $10^3$GeV, will be able to produce micro black holes \cite{DL}.

The Randall-Sundrum (RS) models provide another scenario that explains why gravity is weak relative to the other interactions \cite{RS1,RS2}.
They are based on the metric
\begin{equation}
ds^2=e^{-2kR\phi}g_{\mu\nu}dx^\mu dx^\nu-R^2 d\phi^2,
\end{equation}
where $k$ is a scale on the order of the Planck scale, $g_{\mu\nu}$ is the metric tensor of the 4-dimensional subspace of this 5-dimensional warped spacetime, and $\phi\in[0,\pi]$ is the coordinate for an extra dimension of size $R$.
The Planck energy in the 5-dimensional spacetime, $M_{\text{Pl}(5)}$, is related to the Planck energy of the 4-dimensional spacetime by
\begin{equation}
M_{\text{Pl}}^2=\frac{M_{\text{Pl}(5)}^3}{k}(1-e^{-2\pi kR}).
\end{equation}
In the RS model with a small extra dimension \cite{RS1}, the Standard Model fields are localized on the brane at $\phi=\pi$, so the warp factor $w=e^{\pi kR}=10^{16}$.
The effective 4-dimensional metric tensor $\tilde{g}_{\mu\nu}$ is conformally related to $g_{\mu\nu}$ by
\begin{equation}
g_{\mu\nu}=e^{-2\pi kR}\tilde{g}_{\mu\nu},
\end{equation}
from which a field with the fundamental mass parameter $m_0$ appears to have the physical mass
\begin{equation}
m=e^{-\pi kR}m_0.
\end{equation}
If $kR\sim12$ then $m_0\sim10^{19}\mbox{GeV}$ gives $m\sim10^3\mbox{GeV}$, reproducing the observed hierarchy between the gravitational and electroweak energy scales.

General relativity, which is the current theory of gravitation, has been confirmed by many experimental and observational tests \cite{Will}.
However, this theory has one problematic feature - the appearance of curvature singularities, which are points in spacetime where the density of matter and curvature are infinite and thus the laws of physics break down.
The Einstein-Cartan (EC) theory of gravity naturally extends GR to include matter with intrinsic spin, which produces torsion, providing a more complete account of local gauge invariance with respect to the Poincar\'{e} group.
It is a viable theory of gravity, which differs significantly from GR only at densities of matter much larger than the density of nuclear matter, and thus it passes all the experimental and observational tests of GR.
In this theory, the curvature of the Riemann-Cartan spacetime, represented by the Einstein tensor $G_{ik}$, is related to the matter distribution, represented by the energy-momentum tensor $T_{ik}$ via the first Einstein-Cartan equation \cite{Hehl}:
\begin{eqnarray}
& & G_{ik}=8\pi G\,T_{ik}-(S^l_{\phantom{l}ij}+2S_{(ij)}^{\phantom{(ij)}l})(S^j_{\phantom{j}kl}+2S_{(kl)}^{\phantom{(kl)}j}) \nonumber \\
& & +4S_i S_k+\frac{1}{2}g_{ik}(S^{mjl}+2S^{(jl)m})(S_{ljm}+2S_{(jm)l}) \nonumber \\
& & -2g_{ik}S^j S_j.
\label{EC}
\end{eqnarray}
The torsion tensor $S^j_{\phantom{j}ik}$ is related to the spin tensor $s^{\phantom{ik}j}_{ik}$ via the second Einstein-Cartan equation \cite{Hehl}:
\begin{equation}
S^j_{\phantom{j}ik}-S_i \delta^j_k+S_k \delta^j_i=-4\pi G\,s^{\phantom{ik}j}_{ik},
\end{equation}
where $S_i$ is the torsion vector.
In GR, the torsion tensor vanishes, reducing (\ref{EC}) to the usual Einstein equations.

We recently showed, using the Papapetrou-Nomura-Shirafuji-Hayashi method of deriving the equations of motion for a test body from the conservation laws for the energy-momentum and spin tensors \cite{Pap}, that the EC theory prevents the formation of singularities if matter is composed of Dirac particles, i.e. quarks and leptons (which form all stars) \cite{Niko}.
The presence of torsion implies that a spinor particle in the Riemann-Cartan spacetime cannot be a point or a system of points, otherwise it would contradict the gravitational field equations.
Instead, such a particle is an extended object whose size is determined by the conditions at which torsion introduces significant corrections to the energy-momentum tensor, i.e. when $8\pi G\,T_{ik}$ and the terms after it in (\ref{EC}) are on the same order \cite{Niko}.
For a particle with mass $m$ and spin $s$, its size is on the order of the Cartan radius $r_C$:
\begin{equation}
\frac{m}{r_C^3}\sim G\biggl(\frac{s}{r_C^3}\biggr)^2.
\label{order}
\end{equation}
For the electron, $r_{Ce}\sim10^{-27}$m, which is much larger than its Schwarzschild radius $\sim10^{-57}$m, so the electron (as well as the other fermions) is nonsingular.
The Cartan density for the electron, $\rho_{Ce}\sim m_e/r_{Ce}^3\sim10^{51}\text{kg}\,\text{m}^{-3}$, gives the order of the maximum density of matter composed of quarks and leptons, averting gravitational singularities in the EC theory, even if a black hole forms.
In GR, the size of a spinor particle is on the order of its Schwarzschild radius and such a particle is represented by a singular worldline.
The mass density of a black hole in the EC theory cannot exceed $\rho_{Ce}$.
This condition gives the order of the minimum mass of a black hole in the EC theory $\sim10^{43}\text{GeV}$ \cite{Niko}, while in GR this mass is on the order of the Planck mass.
Therefore if the EC theory is true, the LHC will not be able to produce micro black holes.

The EC theory, as well as the ADD and RS models, are physically very appealing.
The extra dimensions in the ADD and RS models explain why gravity is so weak relative to the other interactions and introduce only one fundamental energy scale, while the torsion of spacetime in the EC theory prevents the formation of singularities from ordinary matter and introduces an effective ultraviolet cutoff in quantum field theory at distances on the order of the Cartan radius of the electron.
Thus combining either the ADD or RS model with EC theory should result in a theory with all the above advantages.

Since the experimental size of the electron is at least $10^3$ times smaller than the radius $R$ of the extra dimensions in the ADD model, the mass and spin densities in (\ref{order}) must be modified, leading to
\begin{equation}
\frac{mR^n}{r_C^{n+3}}\sim G\biggl(\frac{sR^n}{r_C^{n+3}}\biggr)^2.
\label{mod}
\end{equation}
For the electron, $n=2$ gives $r_C\sim10^{-18}\mbox{m}>r_S$, so the electron is nonsingular.
As $n\rightarrow\infty$, $r_S\rightarrow10^{-19}\mbox{m}\sim r_S$.
The presence of torsion in the ADD model implies that the size of the electron must be on the order of $10^{-19}-10^{-18}$m, which is at least $\sim10^3$ times larger than the experimental upper limit on its radius \cite{rad}.
Therefore large extra dimensions and torsion produced by spin are incompatible: the EC theory of gravity excludes large extra dimensions and the ADD model \cite{ADD} of the weak/Planck hierarchy excludes torsion.

In the RS model with a small extra dimension, the electron corresponds to a Dirac field with the fundamental mass $m_{0e}=e^{\pi kR}m_e\sim10^{13}\mbox{GeV}$.
Substituting this mass into (\ref{mod}) with $n=1$ and $R\sim10^{-33}\mbox{m}$ gives $r_{C0e}\sim10^{-33}\mbox{m}$, which is the Cartan radius in the fundamental 5-dimensional theory.
The correspoding Cartan radius observed in the 4-dimensional spacetime is obtained by the conformal scaling,
\begin{equation}
l=e^{\pi kR}l_0,
\end{equation}
which gives $r_{Ce}=e^{\pi kR}r_{C0e}\sim10^{-17}\mbox{m}$.
This result agrees with the Cartan radius obtained from (\ref{mod}) with $m=m_e$ for the ADD model with $n=1$.
This agreement is consistent with the relation between large extra dimensions in \cite{ADD} and exponential determination of the weak/Planck hierarchy \cite{equ}.
The Schwarzschild radius for the electron in the RS model observed in the 4-dimensional spacetime is given by (\ref{Schw}) with the conformal scaling \cite{BH}: $r_{Se}\sim w(GwmR)^{1/2}\sim10^{-22}\mbox{m}$, as in the ADD model with $n=1$.
Thus the electron is nonsingular in the EC+RS model, $r_{Ce}>r_{Se}$.
The presence of torsion in the RS model implies that the size of the electron must be on the order of $10^{-17}\mbox{m}$, which is $\sim10^5$ times larger than the experimental upper limit on its radius \cite{rad}.
Therefore small extra dimensions and torsion produced by spin are also incompatible: the EC theory of gravity excludes small extra dimensions and the RS model \cite{RS1} of the weak/Planck hierarchy excludes torsion.

If $kR=12$ \cite{BH} then the Cartan radius of the electron in the RS model depends on the size of an extra dimension $R$ according to
\begin{equation}
r_{Ce}\sim\xi^{1/4}e^{9\pi\xi}10^{-29}\mbox{m},
\end{equation}
where
\begin{equation}
\xi=\frac{R}{10^{-33}\mbox{m}}.
\end{equation}
If $R$ is larger than $10^{-33}\mbox{m}$ then $r_{Ce}$ is larger than $10^{-17}\mbox{m}$.
For very large $R$, as in \cite{RS2}, $r_{Ce}$ significantly exceeds the experimental upper limit on the radius of the electron \cite{rad}.
Thus the RS model with an infinite fifth dimension \cite{RS2} cannot be valid if spin produces torsion according to the EC theory of gravity.

To conclude, we found that the ADD and RS models of the weak/Planck hierarchy cannot be combined with the EC theory gravity in which spin of matter produces torsion of spacetime.
Any signal showing that extra dimensions exist would indicate that torsion vanishes or at least that spin cannot produce torsion.
Any signal showing that spin produces torsion would indicate that extra dimensions do not exist.
Since including intrinsic spin as another source of the gravitational field and requiring a complete account of local gauge invariance with respect to the Poincar\'{e} group naturally generalize GR into the EC theory, the latter is probably a more fundamental theory of gravity, especially that it also prevents the formation of singularities from matter and introduces an effective ultraviolet cutoff in quantum field theory.
In this case, the ADD and RS models of extra dimensions cannot be viable.


\begin{thebibliography}{}
\bibitem{ADD} N. Arkani-Hamed, S. Dimopoulos, and G. Dvali, Phys. Lett. B {\bf 429}, 263 (1998); I. Antoniadis, N. Arkani-Hamed, S. Dimopoulos, and G. Dvali, Phys. Lett. B {\bf 436}, 257 (1998); N. Arkani-Hamed, S. Dimopoulos, and G. Dvali, Phys. Rev. D {\bf 59}, 086004 (1999); E. G. Floratos and G. K. Leontaris, Phys. Lett. B {\bf 465}, 95 (1999); A. Kehagias and K. Sfetsos, Phys. Lett. B {\bf 472}, 39 (2000).
\bibitem{Long} J. C. Long, H. W. Chan, A. B. Churnside, E. A. Gulbis, M. C. M. Varney, and J. C. Price, Nature {\bf 421}, 922 (2003).
\bibitem{MP} R. C. Myers and M. J. Perry, Ann. Phys. {\bf 172}, 304 (1986).
\bibitem{rad} H. Dehmelt, Phys. Scr. {\bf T22}, 102 (1988).
\bibitem{Wald} R. Geroch and J. Traschen, Phys. Rev. D {\bf 36}, 1017 (1987); R. M. Wald, arXiv:0907.0412.
\bibitem{DL} S. Dimopoulos and G. Landsberg, Phys. Rev. Lett. {\bf 87}, 161602 (2001).
\bibitem{RS1} L. Randall and R. Sundrum, Phys. Rev. Lett. {\bf 83}, 3370 (1999); P. Callin and F. Ravndal, Phys. Rev. D {\bf 72}, 064026 (2005).
\bibitem{RS2} L. Randall and R. Sundrum, Phys. Rev. Lett. {\bf 83}, 4690 (1999).
\bibitem{Will} C. M. Will, {\em Theory and Experiment in Gravitational Physics} (Cambridge Univ. Press, 1992).
\bibitem{Hehl} F. W. Hehl, P. von der Heyde, G. D. Kerlick, and J. M. Nester, Rev. Mod. Phys. {\bf 48}, 393 (1976); E. A. Lord, {\em Tensors, Relativity and Cosmology} (McGraw-Hill, 1976); V. de Sabbata and M. Gasperini, {\em Introduction to Gravitation} (World Scientific, 1986); V. de Sabbata and C. Sivaram, {\em Spin and Torsion in Gravitation} (World Scientific, 1994); F. W. Hehl, J. D. McCrea, E. W. Mielke, and Y. Ne'eman, Phys. Rep. {\bf 258}, 1 (1995); I. L. Shapiro, Phys. Rep. {\bf 357}, 113 (2002); R. T. Hammond, Rep. Prog. Phys. {\bf 65}, 599 (2002); N. J. Pop{\l}awski, arXiv:0911.0334.
\bibitem{Pap} A. Papapetrou, Proc. Roy. Soc. London A {\bf 209}, 248 (1951); K. Nomura, T. Shirafuji, and K. Hayashi, Prog. Theor. Phys. {\bf 86}, 1239 (1991).
\bibitem{Niko} N. J. Pop{\l}awski, arXiv:0910.1181.
\bibitem{equ} N. Arkani-Hamed, S. Dimopoulos, G. Dvali, and N. Kaloper, Phys. Rev. Lett. {\bf 84}, 586 (2000).
\bibitem{BH} L. A. Anchordoqui, H. Goldberg, and A. D. Shapere, Phys. Rev. D {\bf 66}, 024033 (2002); D. Karasik, C. Sahabandu, P. Suranyi, and L. C. R. Wijewardhana, Phys. Rev. D {\bf 69}, 064022 (2004).
\end{thebibliography}
\end{document}